\documentclass[%
 preprint,
groupedaddress,
superscriptaddress,
 amsmath,amssymb,
 aps,
 prx,
 showpacs
]{revtex4-2}

\usepackage{graphicx}
\usepackage{makecell,booktabs}
\usepackage{dcolumn}
\usepackage{bm}
\usepackage{hyperref}
\hypersetup{
    colorlinks=true,
    urlcolor= blue,
    citecolor=blue,
linkcolor= blue}
\usepackage{xcolor}

\begin{document}


\title{Spin and charge interconversion in Dirac semimetal thin films}

\author{Wilson Yanez}
\author{Yongxi Ou}
\author{Run Xiao}

 \affiliation{Department of Physics, Pennsylvania State University, University Park, Pennsylvania 16802, USA}
 \author{Jahyun Koo}
\affiliation{Department of Condensed Matter Physics, Weizmann Institute of Science, Rehovot 7610001, Israel}
\author{Jacob T. Held}
\author{Supriya Ghosh}
\affiliation{%
Department of Chemical Engineering and Materials Science, University of Minnesota, Minneapolis, Minnesota 55455, USA
}%

\author{Jeffrey Rable}
\author{Timothy Pillsbury}
\affiliation{Department of Physics, Pennsylvania State University, University Park, Pennsylvania 16802, USA}
\author{Enrique Gonz\'{a}lez Delgado}
\affiliation{Department of Physics and Electronics, Humacao Campus of the University of Puerto Rico, 100 Tejas Avenue, Humacao, Puerto Rico 00791-4300 }
\author{Kezhou Yang}
\affiliation{Department of Materials Science and Engineering, Pennsylvania State University, University Park, Pennsylvania 16802, USA}
\author{Juan Chamorro}
\affiliation{Department of Chemistry, Johns Hopkins University, Baltimore, MD USA}
 \author{Alexander J. Grutter}
\author{Patrick Quarterman}
\affiliation{NIST Center for Neutron Research, National Institute of Standards and Technology, Gaithersburg, Maryland 20899, USA}
\author{Anthony Richardella}
\affiliation{Department of Physics, Pennsylvania State University, University Park, Pennsylvania 16802, USA}
\author{Abhronil Sengupta}
\affiliation{School of Electrical Engineering and Computer Science, Pennsylvania State University, University Park, Pennsylvania 16802, USA}
\author{Tyrel McQueen}
\affiliation{Department of Chemistry, Johns Hopkins University, Baltimore, MD USA}
\author{Julie A. Borchers}
\affiliation{NIST Center for Neutron Research, National Institute of Standards and Technology, Gaithersburg, Maryland 20899, USA}
\author{K. Andre Mkhoyan}
\affiliation{%
Department of Chemical Engineering and Materials Science, University of Minnesota, Minneapolis, Minnesota 55455, USA
}%
\author{Binghai Yan}
\affiliation{Department of Condensed Matter Physics, Weizmann Institute of Science, Rehovot 7610001, Israel}
\author{Nitin Samarth}
\email{nsamarth@psu.edu}
\affiliation{Department of Physics, Pennsylvania State University, University Park, Pennsylvania 16802, USA}

\date{\today}

\begin{abstract}

We report spin-to-charge and charge-to-spin conversion at room temperature in heterostructure devices that interface an archetypal Dirac semimetal, $\mathrm{Cd_3As_2}$, with a metallic ferromagnet, $\mathrm{Ni_{0.80}Fe_{0.20}}$ (permalloy). The spin-charge interconversion is detected by both spin torque ferromagnetic resonance and ferromagnetic resonance driven spin pumping. Analysis of the symmetric and anti-symmetric components of the mixing voltage in spin torque ferromagnetic resonance and the frequency and power dependence of the spin pumping signal show that the behavior of these processes is consistent with previously reported spin-charge interconversion mechanisms in heavy metals, topological insulators, and Weyl semimetals. We find that the efficiency of spin-charge interconversion in $\mathrm{Cd_3As_2}$/permalloy bilayers can be comparable to that in heavy metals. We discuss the underlying mechanisms by comparing our results with first principles calculations.

\end{abstract}

\maketitle


\section{\label{sec:level1}Introduction}

Spin-to-charge conversion and its Onsager reciprocal charge-to-spin conversion have been extensively studied in several materials with either strong spin-orbit coupling (such as heavy metals, HMs) or with spin momentum ``locking'' (as in the surface states of topological insulators) \cite{PhysRevB.82.214403_Mosendz2010,PhysRevB.88.064414_Vlaminck2013,y1MihaiMiron2010,y2Liu555,y3PhysRevLett.120.097203,y46516040,Tungsten_doi:10.1063/1.4753947,PhysRevLett.117.076601_Hailong2016,PhysRevResearch.1.012014_Hailong2019,Mellnik2014, Kondou2016, Wang_2018}. The latter class of materials motivated the emergence of ``topological spintronics,'' broadly aimed at exploiting the strong spin-momentum correlation in helical topological states for spintronic devices. Theoretical studies provide strong motivation for extending such studies to topological semimetals. For example, in Weyl semimetals, calculations predict a large spin Hall conductivity, the Rashba-Edelstein effect,  and efficient spin-to-charge conversion in magnetic Weyl semimetal-normal metal heterostructures  \cite{PhysRevLett.117.146403_sun, PhysRevB.97.085417_edelstein,PhysRevLett.123.187201_spin2chargeMWSM}. Recent experiments have reported current-induced spin-orbit torques in the Weyl semimetal $\mathrm{WTe_2}$ with efficiencies comparable to HMs \cite{MacNeill2016_Wte2_1,MacNeill_PhysRevB.96.054450}.  In addition, field-free current-induced magnetization switching has been demonstrated at room temperature in $\mathrm{WTe_2/ferromagnet}$ heterostructures with current densities lower than HMs and topological insulators \cite{Shi2019_Wte2_2}. This context motivates the exploration of spin-charge interconversion in a different class of topological semimetals of contemporary interest, the Dirac semimetal (DSM), where only a few studies of spin transport have been reported  \cite{PhysRevLett.124.116802FermiArcs,PhysRevApplied.14.054044FermiArcs2, https://doi.org/10.1002/adma.202000513_PtTe2}. We might anticipate that the interplay between a bulk three-dimensional linear dispersion with non-trivial topology and the spin polarization of surface states could give rise to efficient spin-charge interconversion and thus elicit interest for spintronics \cite{FermiArcTransportCdAs_PhysRevLett.124.116802,Xu1256742_Na3Bi_Arpes, PhysRevB.97.085417_edelstein,Shi2019_Wte2_2,y5PhysRevApplied.9.011002,y6Fukami_2017}. Unlike the intuitive appeal of topological insulators and Weyl semimetals for topological spintronics, however, the spin degeneracy inherent in the topological band structure of a DSM raises a fundamental question: what is the spin Hall conductivity (SHC) of a DSM and could it be relevant for spintronics? 
 
In this paper, we report measurements of spin-charge interconversion at room temperature in bilayers of the archetypal DSM $\mathrm{Cd_3As_2}$ and a conventional metallic ferromagnet, $\mathrm{Ni_{0.80}Fe_{0.20}}$ (permalloy, Py) via spin torque ferromagnetic resonance (ST-FMR) and ferromagnetic resonance driven spin pumping (SP). We also compare our experimental results with first principles calculations of the SHC in $\mathrm{Cd_3As_2}$. In bilayers with an imperfect (oxidized) interface, we find a SHC that is much larger than predicted by theory at the estimated chemical potential and comparable to that of HMs. We attribute the dominant contributions to extrinsic effects. Measurements of bilayers with a clean interface show a smaller SHC, consistent with our theoretical predictions for intrinsic contributions to spin-charge interconversion. 

\section{Sample synthesis and characterization}

We first discuss the synthesis, as well as structural and interfacial characterization, of $\mathrm{Cd_3As_2/Py}$ heterostructures. We used molecular beam epitaxy (MBE) to grow $\mathrm{Cd_3As_2}$ thin films under ultrahigh vacuum conditions (pressure $P < 10^{-7}$ Pa) in a Veeco 930 system. We first grew an intrinsic, relaxed GaSb (111) buffer layer (100 nm thick) on a semi-insulating GaAs(111)B substrate using elemental Ga (5N) and Sb (5N) source materials evaporated from standard effusion cells. The growth conditions used were standard for the III-V semiconductor. The $\mathrm{Cd_3As_2}$ layer was then deposited at a substrate temperature $T_s \sim 200\ ^{\circ}$C. For most of the samples discussed in this paper, the source materials are elemental Cd (5N purity) and As (5N purity) evaporated from conventional effusion cells; the beam equivalent flux ratio of Cd:As was 3:2. Some samples were also grown by evaporating a compound $\mathrm{Cd_3As_2}$ source, also from an effusion cell. Typical growth rates were 0.5 nm/min. During MBE growth, we obtain an unreconstructed streaky reflection high energy electron diffraction (RHEED) pattern (Figs. 1 (a) and 1 (b)) which shows the expected $C_3$ symmetry of the $\mathrm{Cd_3As_2}$ crystal. X-ray diffraction (Fig. 1c) shows peaks corresponding to $\mathrm{Cd_3As_2}$ grown in the $\langle 112 \rangle$ orientation as well as the peaks of the GaSb (111) buffer layer and the GaAs (111) substrate. Typical double crystal rocking curves for diffraction peaks from $\mathrm{Cd_3As_2}$ have widths of $\sim 0.11^{\circ}$. Atomic force microscopy (AFM) (Fig. 1(d)) shows large domains (500 nm lateral islands) with 0.5 nm high steps. These are all signatures of epitaxial growth of reasonable quality $\mathrm{Cd_3As_2}$ thin films, similar to those reported in prior literature \cite{Schumann_APL_2016}. 

The measurement of spin-charge interconversion in $\mathrm{Cd_3As_2}$ requires the synthesis of heterostructures that interface a well-characterized ferromagnet with the DSM. To this end, we transferred the MBE-grown $\mathrm{Cd_3As_2}$ films to a different vacuum chamber for the deposition of Py thin films using an e-beam evaporator. All the spin-charge interconversion data in the main manuscript is taken on samples that involved brief (10-15 minutes) exposure of the $\mathrm{Cd_3As_2}$ film to ambient atmosphere before the subsequent deposition of a Py thin film, followed by 1 nm to 3 nm of either Ta or Al as a capping layer to prevent oxidation. Although technical limitations currently constrain our capability for the routine deposition of Py on $\mathrm{Cd_3As_2}$ without breaking vacuum, we do include some spin-charge interconversion data on a sample grown with a pristine interface (see Appendix D). 

We further characterized the $\mathrm{Cd_3As_2/Py}$ heterostructures using atomic-resolution high-angle annular dark-field (HAADF) scanning transmission electron microscopy (STEM). An example of such data for a heterostructure with brief ambient exposure between MBE growth of $\mathrm{Cd_3As_2}$ and Py deposition is shown in Figs. 1 (e) and 1 (f): the former reveals coherent growth of the entire heterostructure, albeit with the presence of a 1 nm cadmium oxide layer at the $\mathrm{Cd_3As_2/Py}$ interface as confirmed by energy dispersive x-ray (EDX) spectroscopy. More detailed STEM and EDX characterization is available in Appendix D. 
Since proximity-induced magnetism at the interface between the DSM and metallic ferromagnet might play a role in spin-charge interconversion, we also used polarized neutron reflectometry (PNR) measurements to characterize the magnetic behavior at the $\mathrm{Cd_3As_2}$/Py interface. We do not observe any evidence for proximity-induced magnetism in the $\mathrm{Cd_3As_2}$ layer (see Appendix E).
\begin{figure}[]
\includegraphics[width=85mm]{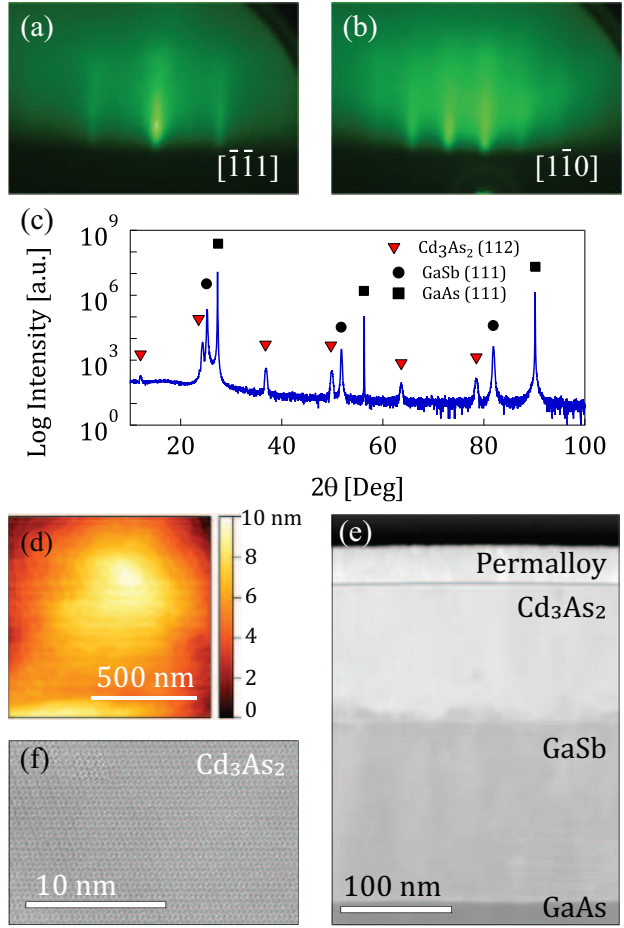}
\caption{\label{fig:1} RHEED pattern of the $\mathrm{Cd_3As_2}$ film along the $[\bar{1}\bar{1}1]$ (a) and $[1\bar{1}0 ]$ (b) crystal directions. (c) X-ray diffraction $2\theta$ scan of $\mathrm{Cd_3As_2}$ films. (d) AFM image of the $\mathrm{Cd_3As_2}$ surface. High-angle annular dark-field TEM image showing the different layers in a complete device stack  (e) and just the $\mathrm{Cd_3As_2}$ layer (f).}
\end{figure}

To confirm the presence of a DSM state in the MBE-grown $\mathrm{Cd_3As_2}$ thin films, we carried out angle resolved photoemission spectroscopy (ARPES) measurements on a few samples that were transferred from the MBE chamber to a local ARPES measurement chamber while maintaining a vacuum environment with pressure $1.3 \lesssim \mu \mathrm{Pa}$. We used the 21 eV helium I$\alpha$ spectral line from a helium plasma lamp isolated via a monochromator as the photon source. The emitted photoelectrons were imaged by a ScientaOmicron DA 30L analyzer with a maximum spectral resolution of 6 meV. Figures 2 (a) and 2 (b) show ARPES measurements from a $\mathrm{Cd_3As_2}$ thin film at room temperature. The spectra are similar to those reported in the prior literature on cleaved bulk crystals of $\mathrm{Cd_3As_2}$ \cite{Liu2014,PhysRevLett.113.027603_ExpReal3DDirac, Yi2014}. The data shows the characteristic linear dispersion in the vicinity of the charge neutral point expected for a DSM. The ARPES data also indicate that the charge neutral point is located 0.2 eV below the chemical potential. This is consistent among several samples with different thicknesses. It is important to note that ARPES cannot probe the surface state of these $\mathrm{Cd_3As_2}$ films once the surface has been capped with Py. Thus, we do not have any direct knowledge of the nature of the electronic states at the $\mathrm{Cd_3As_2}$/Py interface. This is a limitation in all the published studies on topological spintronics thus far. 

We carried out electrical magneto-transport measurements on As-capped $\mathrm{Cd_3As_2}$ thin films at room temperature, revealing a Hall resistance with non-linear dependence on magnetic field, indicative of coexisting electron and hole type of carriers as expected for a chemical potential close to the charge neutral point \cite{liu2015,xiu1, doi:10.1021/acsnano.6b01568}. In the low field limit (the range needed for ST-FMR and SP experiments), the electrical transport is mainly electron type with a carrier density and mobility of around $n= 10^{18} - 10^{19}\ \mathrm{cm}^{-3}$ and $\mu=4000-6000\ \mathrm{cm}^2 \mathrm{V}^{-1} \mathrm{s}^{-1}$.

\begin{figure}[b]
\includegraphics[width=85mm]{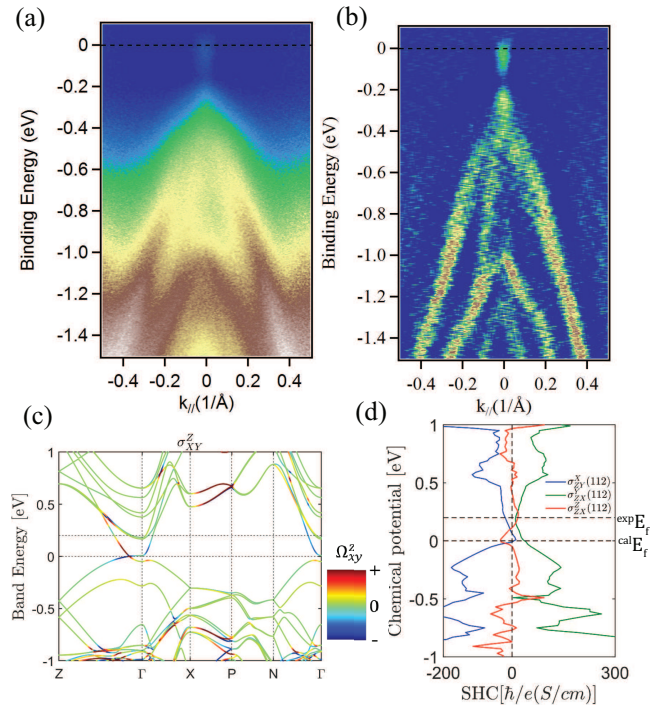}
\caption{\label{fig:2} (a) Electronic band structure of $\mathrm{Cd_3 As_2}$ thin films obtained by ARPES. The measurements are taken at $T = 300$ K along the $M-\Gamma-M$ direction, projected on the (112) surface.  (b) Second derivative of the spectrum shown in (a). (c) Calculated band structure and spin Hall conductivity (SHC) of bulk $\mathrm{Cd_3 As_2}$ The color in the band structure shows the spin Berry curvature ($\Omega_{xy}^z$) with \textit{z} along the [001] direction. (d) Three independent components of SHC with \textit{z} along the [112] direction as a function of chemical potential.}
\end{figure}

\section{Theoretical calculation of spin Hall conductivity}

To obtain a theoretical understanding of the spin-charge interconversion in $\mathrm{Cd_3As_2}$, we performed density-functional theory (DFT) calculations in the framework of the generalized gradient approximation\cite{perdew1996} with a full-potential local-orbital minimum-basis code (FPLO) \cite{koepernik1999full}. Spin-orbit coupling was included. The crystal structure of Cd$_3$As$_2$ belongs to space group $I4_1/acd$ (No.142) and has inversion symmetry \cite{ali2014crystal}. 
From DFT calculations, we projected the Bloch wave function into atomic-like Wannier functions of Cd-5\textit{s} orbital and As-4\textit{p} orbitals. Then we build an accurate tight-binding Hamiltonian($\hat{H}$) to calculate the intrinsic spin Hall conductivity components (Figs. 2(c) and 2(d))~\cite{sinova2015spin}. Full details of the calculation are given in Appendix A.  

In addition, our calculation reveals an out-of-plane spin polarization in the spin current generated by the intrinsic SHC
$\sigma^{z}_{zx} = 16 $ S$\cdot$cm$^{-1}$ ($\hbar/e$), where spin current direction \textit{z} refers to the [112] direction. The non-zero $\sigma^{z}_{zx}$ is induced by the symmetry breaking along the [112] axis. If the intrinsic SHE prevails under certain experimental conditions, we expect that the out-of-plane polarization may play a significant role in the spin-orbit torque. We also find that the SHC has several local maxima around the charge neutral point. This could tentatively explain the variation of the spin-charge interconversion efficiency with changes of chemical potential seen in Ref. \cite{PhysRevLett.124.116802FermiArcs,PhysRevApplied.14.054044FermiArcs2}. It also offers an opportunity to enhance the intrinsic SHC in $\mathrm{Cd_3As_2}$ thin films via chemical doping or electrostatic gating.

\section{Spin torque ferromagnetic resonance}
We now discuss the spin-charge interconversion in $\mathrm{Cd_3As_2}$/Py bilayers via ST-FMR. We have measured a total of 12 devices all showing the interconversion phenomenon. In all these devices except for one, the interface between $\mathrm{Cd_3As_2}$ and Py has some oxidation due to brief ambient exposure. In ST-FMR experiments, we apply a radio frequency (rf) charge current in devices with lateral device dimensions $50\ \mu {\mathrm{m}} \times\ 10\ \mu {\mathrm{m}}$ patterned from bilayers of 12 nm thick $\mathrm{Cd_3As_2}$ and 4 and 6 nm thick Py in the presence of an in-plane external magnetic field. This process excites the magnetization dynamics of the ferromagnetic layer by means of Oersted field and spin current at the interface. The latter results from the rf charge current creating a flow of angular momentum perpendicular to the interface (charge-to-spin conversion) in the DSM (Fig 3(a)). The magnetization dynamics are then probed using the anisotropic magnetoresistance of the ferromagnetic layer which, along with the rf current, produces a DC mixing voltage ($V_{mix}$) in the device. The in-plane and out-of-plane torques can then be extracted by fitting the $V_{mix}$ with a symmetric and antisymmetric Lorentzian function:

\begin{equation}
    V_{mix}= \frac{S \Delta_{Cd_3As_2}^2+A\Delta_{Cd_3As_2}(H-H_{Res})}{\Delta_{Cd_3As_2}^2+(H-H_{Res})^2}.
    \label{eq:vmix}
\end{equation}

Here, $\Delta_{Cd_3As_2}$ is the line width of the absorption spectrum, $H$ is the applied external field, $H_{Res}$ is the field at which we see the resonance and $S(A)$ is the magnitude of the symmetric (antisymmetric) component. The frequency dependence of the absorption spectrum is well understood for thin films using the Kittel equation $f=\frac{\gamma}{2\pi}\sqrt{H_{Res}(H_{Res}+M_{Eff})}$. This can be fitted to our data using the known gyromagnetic ratio of the electron ($\gamma$), giving effective magnetization values of around $M_{Eff}= 560\ \mathrm{kA/m}$ to $640 \  \mathrm{kA/m}$ which is similar to values reported by FMR studies in Py films \cite{Lesne2016_LaO}.

In addition, we use the ratio of the symmetric and antisymmetric components to extract the spin torque efficiency defined as the ratio of the rf spin current ($J_s$) generated in the device and the charge current ($J_c$) applied by an external source  \cite{PhysRevLett.106.036601_luqiao,y8PhysRevB.92.064426}:

\begin{equation}
   \xi=\frac{2e}{\hbar}\frac{J_s}{J_c} = \frac{S}{A} \frac{e\mu_0M_St_{Cd_3As_2}t_{NiFe}}{\hbar}\left[1+\left(\frac{M_{Eff}}{H_{Res}}\right)\right]^{1/2}.
   \label{eq:xi}
\end{equation}

Here, $e$ is the charge of the electron, $\hbar$ is the reduced Planck constant, $\mu_0$ is the permeability of free space, $M_S=560\ \mathrm{kA/m}$ is the saturation magnetization of Py (measured in a Quantum Design superconducting quantum interference device magnetometer), $t_{Cd_3As_2}$ is the thickness of the $\mathrm{Cd_3As_2}$ layer, and $t_{NiFe}$ is the thickness of the ferromagnet.


Figure 3 (b) shows the mixing voltage of a device at different frequencies ranging from 3.5 GHz to 5 GHz.  The spin torque efficiency $\xi$ is extracted from the measured signal by separating it into a symmetric and an antisymmetric Lorentzian component, as shown in Fig. 3 (c). If we assume that the spin current has a bulk origin, we obtain a spin torque efficiency $\xi=0.27$ (0.22) with 6 (4) nm of Py, comparable to values in HM and providing a lower bound on the spin Hall angle due to the less than ideal spin transparency \cite{y8PhysRevB.92.064426}. The angle dependence (Fig. 3(d)) also shows the expected symmetry of the in-plane and out-of-plane torques from, for example, the bulk spin Hall effect (SHE) or Rashba-Edelstein effect \cite{Mellnik2014,PhysRevLett.106.036601_luqiao, y9PhysRevB.94.140414,y10Ou2019}. It also shows no spin polarization in the out-of-plane or current flow directions (see Appendix B). Moreover, the symmetric component of the torque shows the same sign as in Pt and opposite to that in W. The spin torque efficiency can then be combined with the $\mathrm{Cd_3As_2}$ conductivity ($\sigma_{xx}\approx1250\ \mathrm{Scm^{-1}}$) to extract the SHC: 

\begin{equation}
   \sigma_{SH}=\frac{\hbar}{2e}\theta_{SH}\sigma_{xx} \gtrsim 140 \frac{\hbar}{e} \frac{\mathrm{S}}{\mathrm{cm}}
    \label{eq:tsh}
\end{equation}

As a first step to better understand the intrinsic contributions to the SHC of $\mathrm{Cd_3As_2}$, we have studied ST-FMR in a device fabricated from $\mathrm{Cd_3As_2/Py}$ bilayers grown using a full {\it in vacuo} transfer procedure. STEM characterization of such a sample showed a clean $\mathrm{Cd_3As_2/Py}$ interface. The ST-FMR signal is however weaker than in the devices fabricated from bilayers with an oxidized interface. Analysis of the ST-FMR data shows that the spin torque efficiency ($\xi=0.10$) and the SHC ($\sigma_{SH}=63 \frac{\hbar}{e} \frac{S}{cm}$) are closer to the intrinsic theoretical value (see Appendix D), indicating the possibility that the interfacial cadmium oxide layer enhances the extrinsic contributions to the SHC.

\begin{figure}[]
\includegraphics[width=85mm]{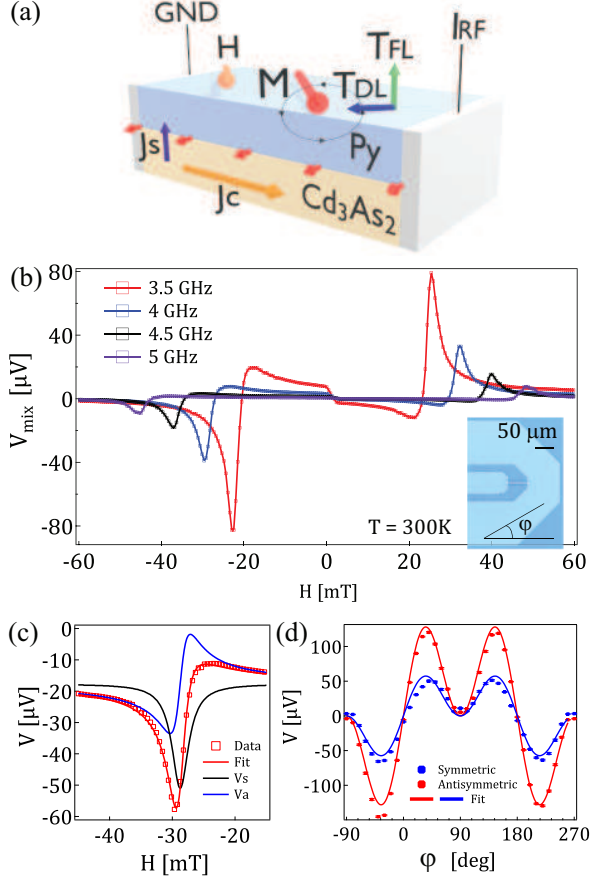}
\caption{\label{fig:3} (a) Illustration of charge-to-spin conversion due to ST-FMR.  (b) Measured ST-FMR mixing voltage signal of a Py(4nm)/$\mathrm{Cd_3As_2}$(12nm) heterostructure with a microwave power of 20 dBm at different frequencies (voltage offset has been removed for clarity). Inset: Optical microscope image of the ST-FMR device. (c) ST-FMR spectra of the same device at 4 GHz and 20 dBm showing the measured data, and the fit needed to extract the symmetric (Vs) and antisymmetric (Va) components of the torque. (d) Angle dependence of the symmetric and antisymmetric components of the torque fitted with $sin(\phi)cos(\phi)^2$}
\end{figure}

\section{Spin pumping}

To complement the ST-FMR measurements of spin-charge interconversion, we also carried out SP measurements on $\mathrm{Cd_3As_2/Py}$ heterostructures. All the data presented here was measured using samples with a partially oxidized interface. In these experiments, we placed the $\mathrm{Cd_3As_2/Py}$ heterostructure in a grounded microstrip transmission line. We studied the FMR of the Py ferromagnet by applying a fixed rf signal in the transmission line while sweeping an external magnetic field (Fig. 4(a)) \cite{PhysRevB.82.214403_Mosendz2010,PhysRevB.88.064414_Vlaminck2013,PhysRevLett.117.076601_Hailong2016,Jamali2015}. The external magnetic field excites the dynamics of the ferromagnet, generating precession of its magnetization which can then be measured as a change in the power absorbed under resonance conditions (Fig. 4(c)) \cite{PhysRevB.82.214403_Mosendz2010,PhysRevLett.117.076601_Hailong2016,Wang_2018}. The frequency dependence of this FMR phenomenon is again characterized by the Kittel equation. As we performed the FMR experiment, we simultaneously measured the voltage generated in the $\mathrm{Cd_3As_2}$ film under resonance conditions (Fig 4(b)). This signal can be decomposed into two contributions. The dominant contribution is linear in power (Fig. 4(d)) and changes sign with magnetic field direction ($V_{SP}$). The secondary contribution is less than $1.5\ \mu V$ and does not change sign under field reversal (see Appendix C)\cite{Seebeck_PhysRevLett.113.196601}. The behavior of $V_{SP}$, which we attribute to spin to charge conversion due to spin pumping from the ferromagnet is similar to the inverse spin Hall effect (ISHE) in HMs and the inverse Rashba-Edelstein effect (IREE) in topological insulators and 2D electron systems in which an electric field ($\vec{E}$) is generated in a direction perpendicular to the spin current ($J_S$) and its spin polarization ($\sigma$), i.e: $\vec{E} \propto \vec{J_{s}}\times\vec{\sigma}$ \cite{PhysRevB.82.214403_Mosendz2010,PhysRevLett.117.076601_Hailong2016,Wang_2018,Lesne2016_LaO}. Again, the sign of the voltage signal matched that of Pt, consistent with our ST-FMR results. 

We have measured SP in several heterostructures with $\mathrm{Cd_3As_2}$ film thickness ranging from 12 nm to 200 nm (measured by STEM and x-ray reflectivity) and a Py thickness of 30 nm. Additionally, as a control experiment, we performed the same SP measurement in a sample with just Py and the GaSb buffer layer; we did not detect a voltage signal under resonance conditions. Thus, we are sure that the spin-to-charge conversion is due to the presence of the $\mathrm{Cd_3As_2}$ layer. Finally, we use the broadening of the linewidth of the absorption spectra in the $\mathrm{Cd_3As_2}$ samples ($\Delta H_{Cd_3As_2}$) and the $\mathrm{GaSb}$ samples ($\Delta H_{GaSb}$) to compare the spin mixing conductance ($g_{\uparrow\downarrow}$) in the two cases \cite{PhysRevB.82.214403_Mosendz2010,PhysRevLett.117.076601_Hailong2016,y7PhysRevLett.88.117601}:  

\begin{equation}
   g_{\uparrow\downarrow}=\frac{2\pi\sqrt{3} M_S \gamma t_{NiFe}}{g\mu_B\mathrm{w}}(\Delta H_{Cd_3As_2}-\Delta H_{GaSb}).
   \label{eq:g}
\end{equation}
Here, $g$ is the Land\'{e} factor, $\mu_B$ is the Bohr magneton, and $w$ is the width of our samples (ranging between 5 mm and 12 mm).
We obtain a spin mixing conductance in the range between $1.0 \ \times \ 10^{18} \ \mathrm{m}^{-2}$ to $3.5\ \times \ 10^{18} \ \mathrm{m}^{-2}$. The difference in spin mixing conductance amongst samples might be due to variations in the $\mathrm{Py/Cd_3As_2}$ interface during fabrication. 
We can further compute the spin current pumped into the $\mathrm{Cd_3As_2}$ layer using equation \cite{PhysRevLett.117.076601_Hailong2016,PhysRevB.82.214403_Mosendz2010}:
\begin{equation}
    J_S=\frac{2e}{\hbar}\frac{g_{\uparrow \downarrow}h_{RF}^2\hbar\omega^2\left[\gamma 4\pi M_S+\sqrt{(\gamma 4\pi M_S)^2+4\omega^2}\right]}{2\pi(\Delta H_{Cd_3As_2})^2 \left[(\gamma 4\pi M_S)^2+4\omega^2\right]}.
\end{equation}

Here, $\omega$ is the resonance frequency and $h_{rf}$ is the magnetic field generated by the transmission line which has been calibrated using the ISHE of Pt and the IREE of $\mathrm{Bi_2Se_3}$. We obtain values of $J_S$ that range between $0.5\ \times 10^{4}\ \mathrm{Am^{-2}}$ to $3.4\ \times 10^{4}\ \mathrm{Am^{-2}}$. 


Finally, we compute the efficiency of the spin-to-charge conversion by calculating the charge current generated in the film by normalizing the measured voltage signal by the sample width (w) and its resistance ($R$). Since DSMs can have both surface and bulk states \cite{Xu1256742_Na3Bi_Arpes}, we tentatively describe the efficiency of the conversion in terms of bulk spin Hall angle ($\theta_{SH}$) with spin diffusion length ($\lambda_{SD}$) and surface inverse Rashba-Edelstein effect with effective length ($\lambda_{IREE}$) \cite{PhysRevB.82.214403_Mosendz2010,PhysRevLett.117.076601_Hailong2016,PhysRevResearch.1.012014_Hailong2019}. 




\begin{equation}
    \frac{Vsp}{\mathrm{w}R}=-\left[\theta_{SH} \lambda_{SD} \tanh \left(\frac{t_{Cd_3As_2}}{2\lambda_{SD}}  \right)+\lambda_{IREE}\right] J_S
    \label{eq:vsp}
\end{equation}



Large variations in sample fabrication (differences in roughness, carrier density, and $\mathrm{Py/Cd_3As_2}$ interface) prevent reliable fits to equation \ref{eq:vsp}. Nevertheless, if we assume just a bulk contribution with a spin diffusion length in $\mathrm{Cd_3As_2}$ that is smaller than the thickness of our thicker samples (200 nm), we can approximate $\theta_{SH}\lambda_{SD} \approx 0.1\ \mathrm{nm} - 1.2\ \mathrm{nm}$. This large range might be due to differences in the interface among samples.

\begin{figure}[]
\includegraphics[width=85mm]{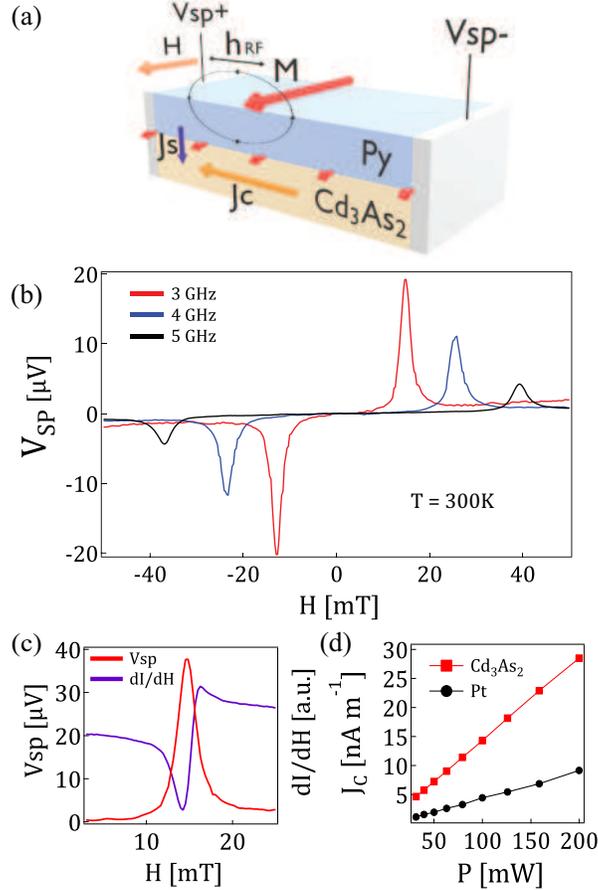}
\caption{\label{fig:4} (a) Illustration of spin-to-charge conversion due to SP. (b) Voltage signal measured in the $\mathrm{Cd_3As_2}$(40 nm)/Py(30 nm) heterostructure at room temperature with an applied microwave power of 20 dBm. (c) SP voltage signal $(V_{SP})$ and FMR absorption spectra $(dI/dH)$ measured at 3 GHz and 23 dBm. (d) Charge current ($J_C$) as a function of power for $\mathrm{Cd_3As_2}$(40 nm)/Py(30 nm) and Py(30 nm)/Pt(20 nm) heterostructures showing linear behavior. }
\end{figure}

\section{Discussion}

Our measurements of spin-charge interconversion at room temperature in $\mathrm{Py/Cd_3As_2}$ heterostructures reveals $\xi\approx 0.22-0.27$, $\sigma_{SH} \gtrsim 140\ h/e\ S\mathrm{cm}^{-1}$ and $\theta_{SH}\lambda_{SD} \approx 0.1\ \mathrm{nm}-1.2\ \mathrm{nm}$. The spin torque efficiency is comparable to the reported values of HMs such as $\beta$-W and Pt \cite{PhysRevB.82.214403_Mosendz2010,y8PhysRevB.92.064426,Wang_2018,Tungsten_doi:10.1063/1.4753947}. Differences in the interface during fabrication makes it difficult to reliably compare the results in both experiments. Nevertheless, this seems to indicate short ($\lesssim 5.5\ \mathrm{nm}$) spin diffusion lengths in $\mathrm{Cd_3As_2}$. We expect that further improvements will allow us to do a systematic thickness dependence study of this phenomenon to determine the SHC and the spin diffusion length more accurately. 



The SHE can be theoretically classified into intrinsic and extrinsic contributions. As discussed earlier, the intrinsic SHC, which is determined by the spin Berry curvature in the wave function, can be calculated from the first-principles band structure (see Appendix A). Consider $x,y$ in the (112) crystallographic plane and the spin current along $z$, as in the experimental setup. We obtain the related SHC $\sigma^{y}_{zx} = 10$ S$\cdot$cm$^{-1}$ ($\hbar/e$). This is about one order of magnitude smaller than the experimental value $\sim 140$ S$\cdot$cm$^{-1}$ ($\hbar/e$) measured in samples with an imperfect oxidized interface. On the other hand, preliminary measurements of a clean interface between the DSM and FM reveal a SHC more consistent with theory. This implies that the SHE in the samples with an imperfect interface may be dominated by extrinsic effects. These extrinsic contributions can have both surface and bulk origins. There are trivial surface states and DSM Fermi arc states. They can generate spin polarization and induce spin diffusion into the NiFe side when current flows on the surface. Because Fermi arcs originate from the quantum spin Hall edge states, their SHE contribution is equivalent to the intrinsic bulk SHC due to the bulk-boundary correspondence. The extrinsic bulk contribution is related to the spin-scattering and strong spin-orbit coupling in the sample. 

In summary, we have carried out systematic measurements of spin-charge interconversion in a topological semimetal, $\mathrm{Cd_3As_2}$, revealing the importance of large extrinsic contributions to the spin Hall conductivity that can arise at an imperfect interface. These results show that imperfect interfaces can sometimes be useful for enhancing spin-charge conversion efficiency. Our results also raise caution about the interpretation of spin transport measurements in $\mathrm{Cd_3As_2}$ that attribute observations to surface Fermi arcs.

\begin{acknowledgments}

The principal support for this project was provided by SMART, one of seven centers of nCORE, a Semiconductor Research Corporation program, sponsored by the National Institute of Standards and Technology (NIST). This supported the synthesis and standard characterization of $\mathrm{Py/Cd_3As_2}$ heterostructures as well as spin-charge interconversion measurements (WY,NS) and their characterization using STEM (JH, SG, AM). Additional support for materials synthesis was provided by the Institute for Quantum Matter under DOE EFRC grant DE-SC0019331 (RX, JC, NS, TM).  The Penn State Two-Dimensional Crystal Consortium-Materials Innovation Platform (2DCC-MIP) under NSF Grant No. DMR-1539916 provided support for ARPES measurements (YO, TP, AR, NS). The magnetometry measurements were carried out by JR, supported by a grant from the University of Chicago. Part of this work was carried out in the College of Science and Engineering Characterization Facility, University of Minnesota, which has received capital equipment funding from the National Science Foundation through the UMN MRSEC under Award Number DMR-2011401 (JH, SG, AM). EG acknowledges support for an undergraduate summer internship from the Office of Graduate Educational Equity Programs and Eberly College of Science at the Pennsylvania State University. Certain commercial equipment, instruments, or materials (or suppliers, or software, ...) are identified in this paper to foster understanding. Such identification does not imply recommendation or endorsement by the National Institute of Standards and Technology, nor does it imply that the materials or equipment identified are necessarily the best available for the purpose.

\end{acknowledgments}

\newpage

\appendix 

\section{Intrinsic spin Hall effect calculations}

The spin current J$_{i}^{Sk}$ generated by electric field $\overrightarrow{E}$ via spin Hall conductivity (J$_{i}^{Sk}=\sigma_{i j}^k E_j$), where J$_{i}^{Sk}$ flows along $i$ direction when spin polarization along $k$ direction. We evaluate the spin Hall conductivity (SHC) ($\sigma^{k}_{i j}$) by the Kubo-formula approach in the linear response scheme\cite{sinova2015spin},
 \begin{equation}
  \label{sigma}
    \sigma^k_{i j} = \frac{e}{\hbar}\sum_{n}\int_{B Z} \frac {d^3\textbf{k}}{(2\pi)^3} f_n(\textbf{k})\Omega^{k}_{n,i j} (\textbf{k})
 \end{equation}
    
 \begin{equation}
    \label{omega}
    \Omega^{k}_{n,i j} (\textbf{k})=2i\hbar^2\sum_{m\ne{n}} \frac{\langle u_n(\textbf{k})|\hat{j}_{i}^k | u_m(\textbf{k}) \rangle \langle u_m(\textbf{k})|\hat{v_j} | u_n(\textbf{k})\rangle} {(E_n(\textbf{k})-E_m(\textbf{k}))^2}
 \end{equation}

Here, $\epsilon_n$ is the eigenvalue of the $|n\rangle$ eigenstate, and $\hat{v}_i=\frac{d\hat{H}}{\hbar dk_i}$ ($i=x,y,z$)  is the velocity operator, $f_n$ is the Fermi-Dirac distribution. $\hat{j}_i^j$ is the spin current operator which is related to the velocity operator($\hat{v}_i$) and spin operator($\hat{s}_j$) as $\hat{j}_i^j=[\hat{v}_i,\hat{s}_j]$
A $k$-point of grid of $200\times200\times200$ is used for the numerical integration in Equation~\ref{sigma}. 

We set the $x,y,z$ axes as the crystallographic $a,b,c$ axes, respectively. According to the space group symmetry and time-reversal symmetry, the $\sigma^{k}_{i j}$ tensor has only three independent matrix elements, $\sigma^{z}_{xy}$, $\sigma^{y}_{zx}$ and $\sigma^{x}_{yz}$, as shown in Table 1. SHC depends sensitively on the chemical potential, as shown in Fig. 5. 

The DFT band structure is shown in Fig. 5. The Dirac point appears at $k_z = \pm k_D$ between $\Gamma$ and $Z$, well consistent with previous work.
The Dirac bands contribute mainly to the $\sigma^{z}_{xy}$ component. This is consistent with the fact that the $k_z =0 $ plane between two Dirac points is a quantum spin Hall insulator. If we approximate all $k_z$ planes between two Dirac points are quantum spin Hall states, then we can estimate the 3D SHC by,
\begin{equation}
    \sigma^{z}_{xy} = \frac{k_D}{k_Z}(\frac{1}{c/2}) \frac{2e^2}{h}  (\frac{\hbar/2}{e})
\end{equation}
where $k_Z$ is the $\Gamma-Z$ distance in the Brillouin zone, $\frac{2e^2}{h}$ is the conductance quantum, $c$ is the lattice parameter of the tetragonal unitcell, and $\frac{\hbar/2}{e}$ converts the dimension from the charge current to the spin current. By taking $\frac{k_D}{k_Z} = 14 ~\%$ from the band structure and lattice parameter $ c=25.4 $ \AA, we obtain  $\sigma^{z}_{xy} \approx 40 $ S$\cdot$cm$^{-1}$ ($\hbar/e$). This is actually very close to the numerical value in Table 1.

When rotating the $z$ axis to the [112] crystallographic axis, which is approximately the [111] direction in the Cartesian's coordinate. We can transfrom the SHC tensor by the unitary rotation \cite{Seemann2015},
\begin{equation}
    \label{rotation}
    {\sigma_{[112]}}_{i j}^{k}=\sum_{l,m,n}D_{i l}D_{j m}D_{kn} {\sigma_{[001]}}_{l m}^{n}
\end{equation}
where $D_{l m}$ is the rotation matrix element that rotates the $z$ axis to the [112] crystallographic axis. The transformed SHC matrix is shown in Tables 1 and 2. 

When $x,y$ align in the (112) crystallographic plane, the new $x,y,z$ axes are not high-symmetry lines as those before. Therefore, new matrix elements emerge in the SHC. When charge flows in the $x,y$ plane and spin flows along $z$ as the experiment setup, $\sigma^{y}_{zx},\sigma^{z}_{zx},  \sigma^{x}_{zy}$ are the relevant SHC matrix elements. Here, $\sigma^{y}_{zx} = 10$ S$\cdot$cm$^{-1}$ ($\hbar/e$) is about one order of magnitude smaller than the experimental SHC, 140 S$\cdot$cm$^{-1}$ ($\hbar/e$). This suggests that the main contributions to the observed SHE are related to extrinsic effects. 
In addition, $\sigma^{z}_{zx}$ indicates the existence of an out-of-plane spin component in the spin current. Suppose the intrinsic SHE prevails under certain experimental condition, we expect that the out-of-plane polarization may play a significant role in the spin-orbit torque. 


\begin{center}
\includegraphics[width=15cm]{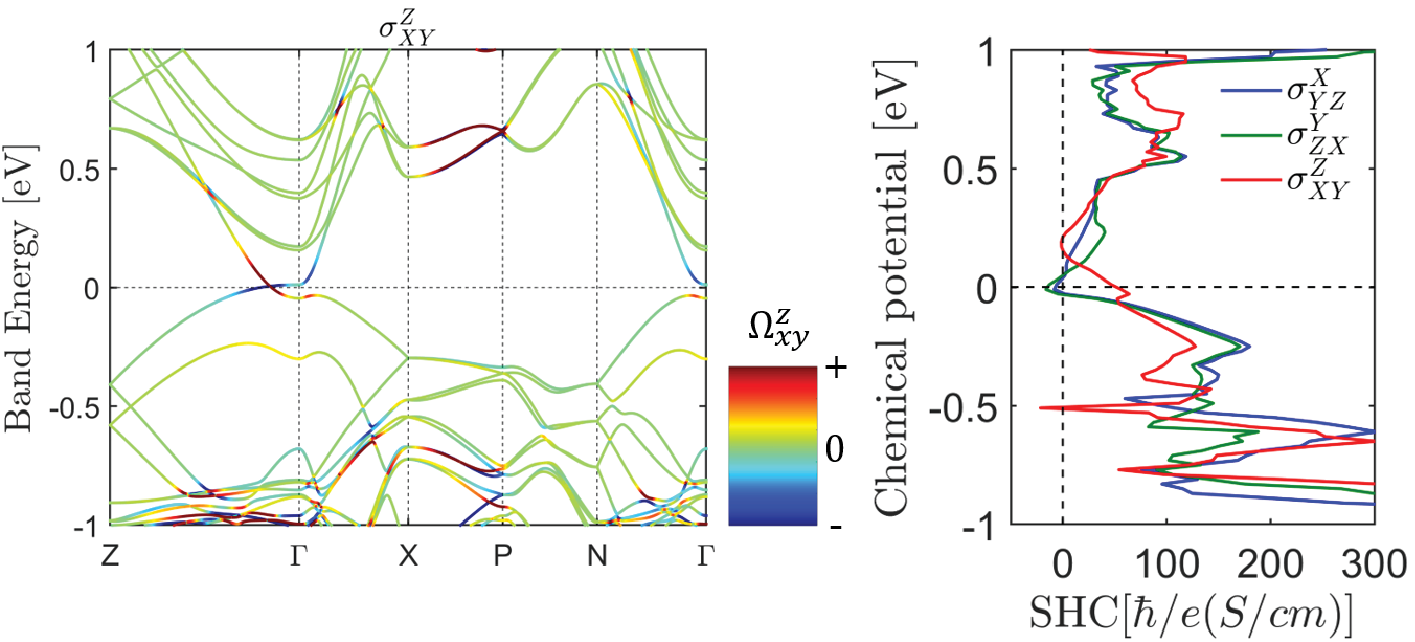}
\end{center}
\noindent {\bf Fig. 5.} Band structure, spin Berry curvature and spin Hall conductivity along the (001) direction. Calculated band structure and spin hall conductivity of bulk Cd$_3$As$_2$. The color in the band structure shows the spin Berry curvature ($\Omega_{xy}^z$). The right panel presents three independent values of spin Hall conductivity calculated in the $\langle 001 \rangle$ direction as a function of chemical potential. 
\label{FigureS1_band}


\newsavebox\sigmaxa
\begin{lrbox}{\sigmaxa}
 $\left(\begin{array}{ccc}0 & 0 & 0 \\0 & 0 & -\sigma^{y}_{xz} \\0 & -\sigma^{y}_{zx}& 0\end{array}\right)$
\end{lrbox}
\newsavebox\sigmaxb
\begin{lrbox}{\sigmaxb}
   $\left(\begin{array}{ccc}0 & 0 & 0 \\0 & 0 & -1 \\0 & 8 & 0\end{array}\right)$
\end{lrbox}
\newsavebox\sigmaxc
\begin{lrbox}{\sigmaxc}
  $ \left(\begin{array}{ccc}0 & 0 & 0 \\0 & -3 & -6 \\0 & 3 & 3\end{array}\right)$
\end{lrbox}

\newsavebox\sigmaya
\begin{lrbox}{\sigmaya}
   $\left(\begin{array}{ccc}0 & 0 & \sigma^{y}_{xz}\\0 & 0 & 0 \\ \sigma^{y}_{zx} & 0 & 0\end{array}\right)$
\end{lrbox}

\newsavebox\sigmayb
\begin{lrbox}{\sigmayb}
   $\left(\begin{array}{ccc}0 & 0 & 1 \\0 & 0 & 0 \\-8 & 0 & 0\end{array}\right)$
\end{lrbox}

\newsavebox\sigmayc
\begin{lrbox}{\sigmayc}
   $\left(\begin{array}{ccc}0 & -19 & -26 \\23 & 0 & 0 \\24 & 0 & 0\end{array}\right)$
\end{lrbox}

\newsavebox\sigmaza
\begin{lrbox}{\sigmaza}
   $\left(\begin{array}{ccc}0 & \sigma^{z}_{xy}  & 0 \\-\sigma^{z}_{xy}  & 0 & 0 \\0 & 0 & 0\end{array}\right)$
\end{lrbox}

\newsavebox\sigmazb
\begin{lrbox}{\sigmazb}
   $\left(\begin{array}{ccc}0 & 40 & 0 \\-40 & 0 & 0 \\0 & 0 & 0\end{array}\right)$
\end{lrbox}

\newsavebox\sigmazc
\begin{lrbox}{\sigmazc}
   $\left(\begin{array}{ccc}0 & 13 & 19 \\-8 & 0 & 0 \\-23 & 0 & 0\end{array}\right)$
\end{lrbox}

\begin{table} [h!]
 \centering
 \caption{Calculated spin Hall conductivity of Cd$_3$As$_2$ at the charge neutral point. The SHC is in the unit of ($\hbar$/e)(S/cm).}
 \resizebox{0.8\textwidth}{!}{%

    \begin{tabular}{cccc}
     & $\sigma^x$ & $\sigma^y$ & $\sigma^z$
       \\ \midrule\midrule 
        \makecell{Magnetic Laue group \\4/mmm z[001]} & \usebox{\sigmaxa} & \usebox{\sigmaya} & \usebox{\sigmaza}\\ \\

    \makecell{Calculation \\ z[001]} & \usebox{\sigmaxb}  & \usebox{\sigmayb} & \usebox{\sigmazb}\\ \\
    
      \makecell{Calculation \\ z[112]} & \usebox{\sigmaxc} & \usebox{\sigmayc} & \usebox{\sigmazc}\\ \midrule\midrule
    \end{tabular}%
}
  \end{table}

\newsavebox\sigmaxaB
\begin{lrbox}{\sigmaxaB}
 $\left(\begin{array}{ccc}0 & 0 & 0 \\0 & 0 & -\sigma^{y}_{xz} \\0 & -\sigma^{y}_{zx}& 0\end{array}\right)$
\end{lrbox}
\newsavebox\sigmaxbB
\begin{lrbox}{\sigmaxbB}
   $\left(\begin{array}{ccc}0 & 0 & 0 \\0 & 0 & 12 \\0 & -32 & 0\end{array}\right)$
\end{lrbox}
\newsavebox\sigmaxcB
\begin{lrbox}{\sigmaxcB}
  $ \left(\begin{array}{ccc}0 & 0 & 0 \\0 & 10 & 25 \\0 & -19 & -10\end{array}\right)$
\end{lrbox}

\newsavebox\sigmayaB
\begin{lrbox}{\sigmayaB}
   $\left(\begin{array}{ccc}0 & 0 & \sigma^{y}_{xz}\\0 & 0 & 0 \\ \sigma^{y}_{zx} & 0 & 0\end{array}\right)$
\end{lrbox}

\newsavebox\sigmaybB
\begin{lrbox}{\sigmaybB}
   $\left(\begin{array}{ccc}0 & 0 & -12 \\0 & 0 & 0 \\32 & 0 & 0\end{array}\right)$
\end{lrbox}

\newsavebox\sigmaycB
\begin{lrbox}{\sigmaycB}
   $\left(\begin{array}{ccc}0 & 6 & -3 \\-15 & 0 & 0 \\10 & 0 & 0\end{array}\right)$
\end{lrbox}

\newsavebox\sigmazaB
\begin{lrbox}{\sigmazaB}
   $\left(\begin{array}{ccc}0 & \sigma^{z}_{xy}  & 0 \\-\sigma^{z}_{xy}  & 0 & 0 \\0 & 0 & 0\end{array}\right)$
\end{lrbox}

\newsavebox\sigmazbB
\begin{lrbox}{\sigmazbB}
   $\left(\begin{array}{ccc}0 & -1 & 0 \\1 & 0 & 0 \\0 & 0 & 0\end{array}\right)$
\end{lrbox}

\newsavebox\sigmazcB
\begin{lrbox}{\sigmazcB}
   $\left(\begin{array}{ccc}0 & 8 & -6 \\-20 & 0 & 0 \\16 & 0 & 0\end{array}\right)$
\end{lrbox}

\begin{table} [h!]
 \centering
 \caption{Calculated spin Hall conductivity of Cd$_3$As$_2$ at the experimental Fermi level (0.2 eV above the charge neutral point). The SHC is in the unit of ($\hbar$/e)(S/cm).}
 \resizebox{0.8\textwidth}{!}{%

    \begin{tabular}{cccc}
     & $\sigma^x$ & $\sigma^y$ & $\sigma^z$
       \\ \midrule\midrule 
        \makecell{Magnetic Laue group \\4/mmm z[001]} & \usebox{\sigmaxaB} & \usebox{\sigmayaB} & \usebox{\sigmazaB}\\ \\

    \makecell{Calculation \\ z[001]} & \usebox{\sigmaxbB}  & \usebox{\sigmaybB} & \usebox{\sigmazbB}\\ \\
    
      \makecell{Calculation \\ z[112]} & \usebox{\sigmaxcB} & \usebox{\sigmaycB} & \usebox{\sigmazcB}\\ \midrule\midrule
    \end{tabular}%
}
  \end{table}

\label{Table_2}



\section{Analysis of spin polarization in ST-FMR measurements}

Following the procedure sketched in \cite{y10Ou2019}, we can study the direction of the spin polarization of the spin current in $\mathrm{Cd_3As_2}$ by performing an angle dependent ST-FMR measurement.
Different polarizations will contribute to different functional forms of the spin current (Fig. 6 (a)):
\begin{equation}
    \label{polarization}
    V_{Sx} \propto sin(\phi)sin(2\phi),\ 
V_{Sy} \propto cos(\phi)sin(2\phi),
V_{Sz} \propto sin(2\phi)
\end{equation}

We fit the symmetric and antisymmetric ST-FMR data with a spin polarization along the y axis and along all three axes (Fig. 6 (c) and (d)). The two fits do not show any significant differences. In the latter case, the amplitudes of the voltage signal with spin polarization along the $x$ and $z$ axis ($V_{Sx}$ and $V_{Sz}$) are less than 5$\%$ the value with spin polarization along the $y$ axis ($V_{Sy}$). Thus, we are confident that the direction of the spin polarization of the measured torque is in the plane of the sample and transverse to the current direction. 
\begin{center}
\includegraphics[width=10cm]{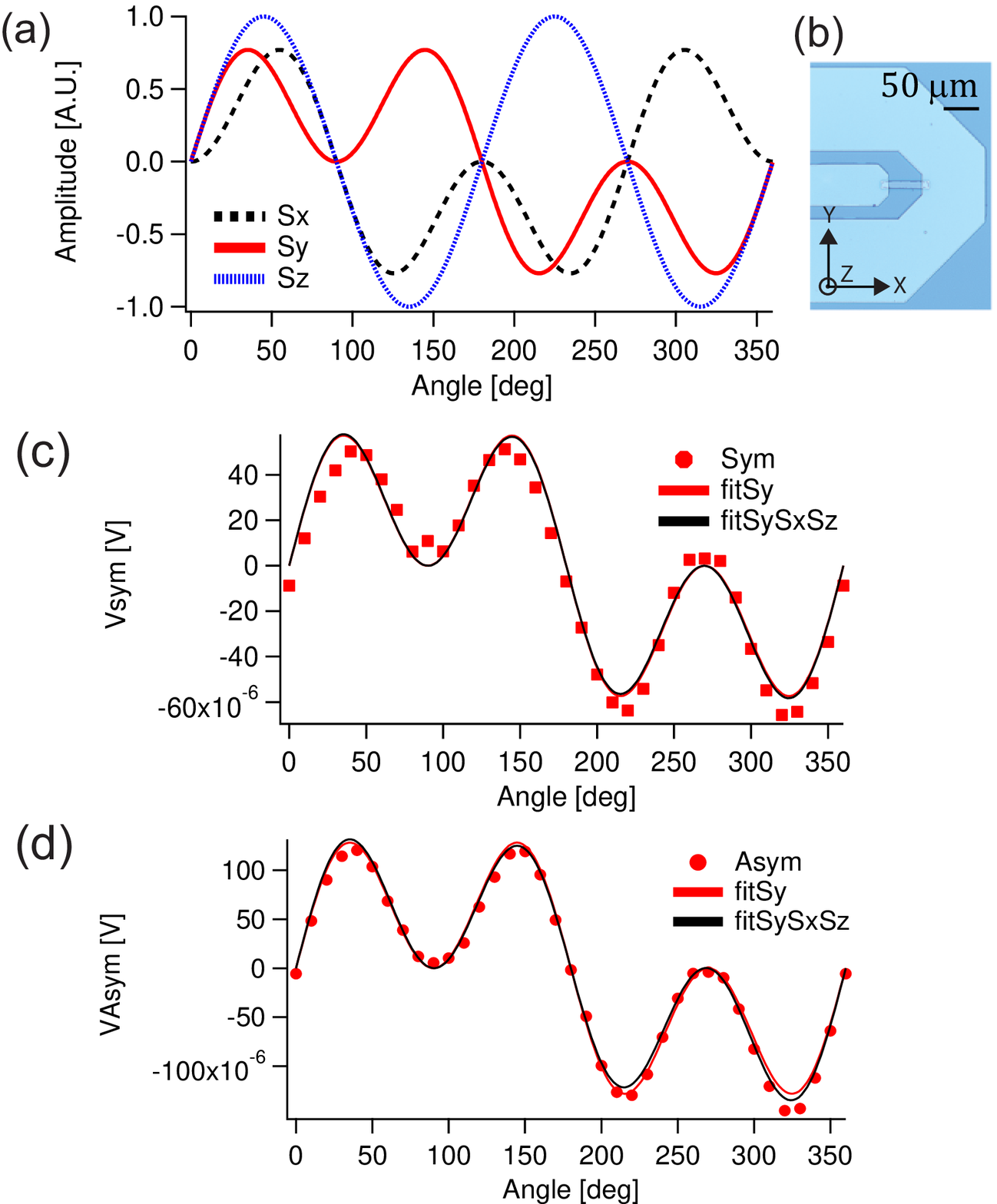}\\
\end{center}
\noindent {\bf Fig. 6.} (a) Theoretical angle dependent functional form of the ST-FMR signal with spin polarization along different axis. (b) Image of one of our samples showing the coordinate axis. 
Angle dependence of the symmetric (b) and antisymmetric components of the torque fitted with spin polarization along the y axis (red) and all axes (black).


\section{Symmetric and antisymmetric components of the spin pumping signal}

As mentioned in the main text, the voltage signal measured in our samples (Fig. 7(a)) can be decomposed in an anstisymmetric signal that changes sign under field reversal ($V_{asym}$ shown in Fig. 7(b)) and a symmetric one that does not ($V_{sym}$ shown in Fig. 7(c)). When we increased the power in the transmission line, we found that $V_{asym}$ increases linearly as expected from ferromagnetic driven spin pumping into the $\mathrm{Cd_3As_2}$ layer from the $\mathrm{NiFe}$ ferromagnet. The symmetric signal has an amplitude of less than $4\%$ of the asymmetric one and is non-linear with power. This is consistent with a possible Seebeck effect induced by the microstrip not being perfectly centered over the sample \cite{Seebeck_PhysRevLett.113.196601}. 

\begin{center}
\includegraphics[width=10cm]{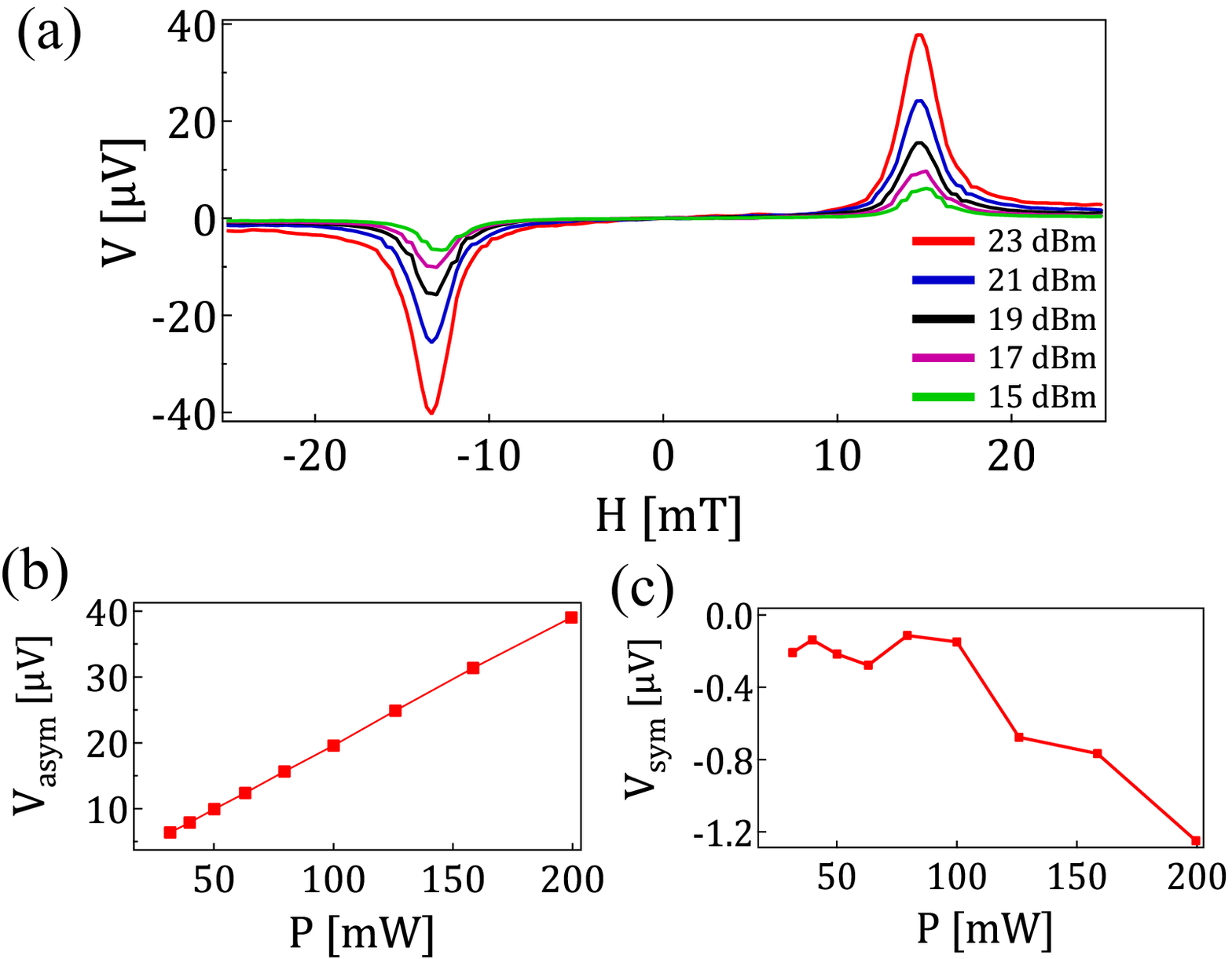}\\
\end{center}

\noindent {\bf Fig. 7.} (a)  Voltage signal measured in a $\mathrm{Py(30\  nm)/Cd_3As_2(40\ nm)}$ heterostructure as we sweep the magnetic field. Magnitude of the antisymmetric (b) and symmetric (c) components of the voltage signal at resonance condition as a function of power.

\section{Study of the Cd3As2/Py interface}

In the samples presented in the main text, atomic resolution high-angle annular dark-field (HAADF) scanning transmission electron microscopy (STEM) and energy dispersive X-ray spectroscopy (EDX) of the $\mathrm{Cd_3As_2/Py}$ heterostructures show a 1 nm to 1.5 nm thick amorphous CdOx layer at the interface (Fig. 8). This probably forms during the ambient condition transfer of samples from the MBE chamber to the metal evaporation chamber.

\begin{center}
\includegraphics[width=15cm]{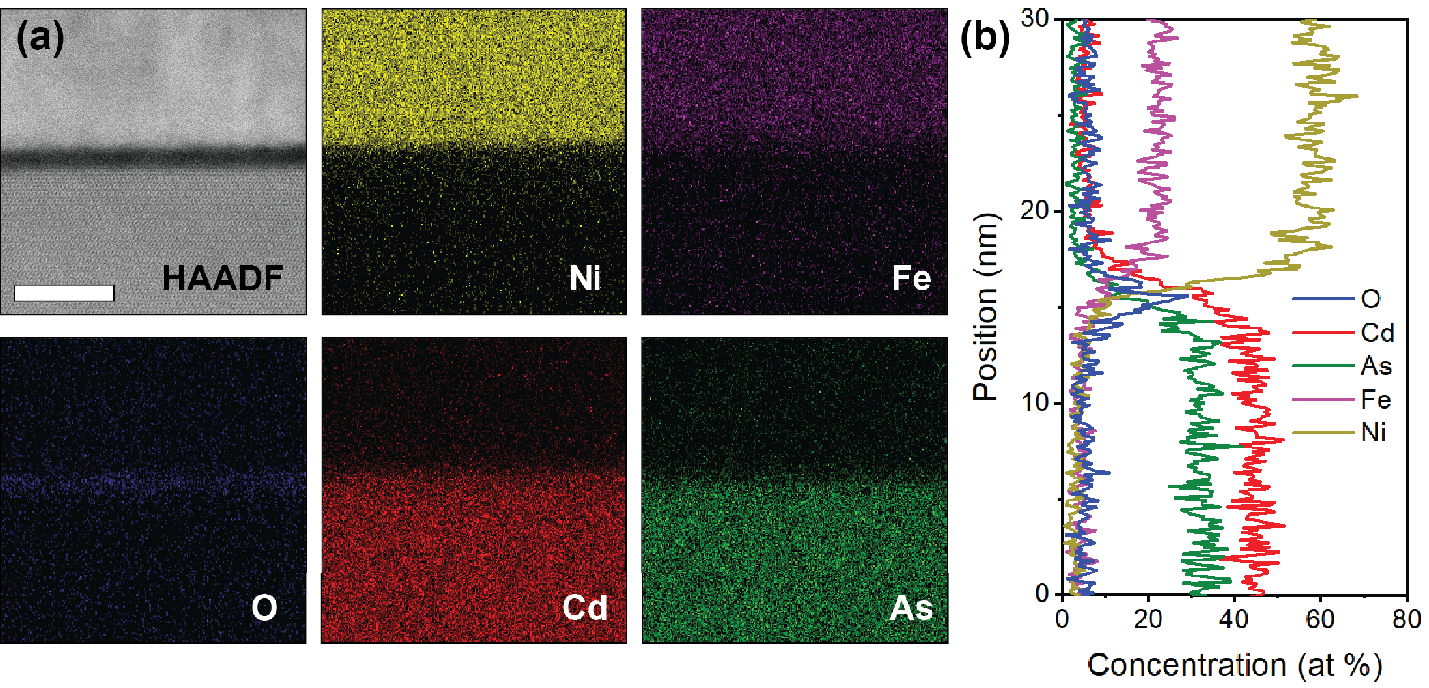}\\
\end{center}

\noindent {\bf Fig. 8.} STEM-EDX elemental map of the $\mathrm{Cd_3As_2/Py}$ interface. (a) HAADF-STEM image and elemental maps of Ni, Fe, O, Cd, and As. Scale bar is 10 nm. (b) Concentration of the elements in (a) across the interface. Note the $\sim1$ nm amorphous layer of $\mathrm{CdO_x}$ at the interface

To avoid the oxidation effect at the $\mathrm{Cd_3As_2/Py}$ interface, we have fabricated ST-FMR devices using a full in vacuum transfer procedure with a vacuum suitcase. Differences in sample holder configurations that allowed the compatibility between chambers seemed to increase the roughness of the $\mathrm{Cd_3As_2}$ layer which increased the noise level in our measurements. Nevertheless as seen in Fig. 9, we have successfully been able to remove the oxide layer at the interface. Moreover, we have been able to measure an ST-FMR signal that clearly shows resonance up to 8 GHz (Fig. 10). This allowed us to compute the spin torque efficiency $\xi=0.10$ and SHC $\sigma_{SH} \gtrsim 63\frac{\hbar}{e}\frac{S}{cm}$ of this device. This experimental value is closer to the intrinsic theoretical one and seems to indicate the possibility that the $\mathrm{CdO_x}$ layer enhances the extrinsic spin Hall effect. Furthermore, the data corroborates that the measured signal is due to charge to spin conversion in the $\mathrm{Cd_3As_2/Py}$ interface as presented in the main text. 

\begin{center}
\includegraphics[width=12cm]{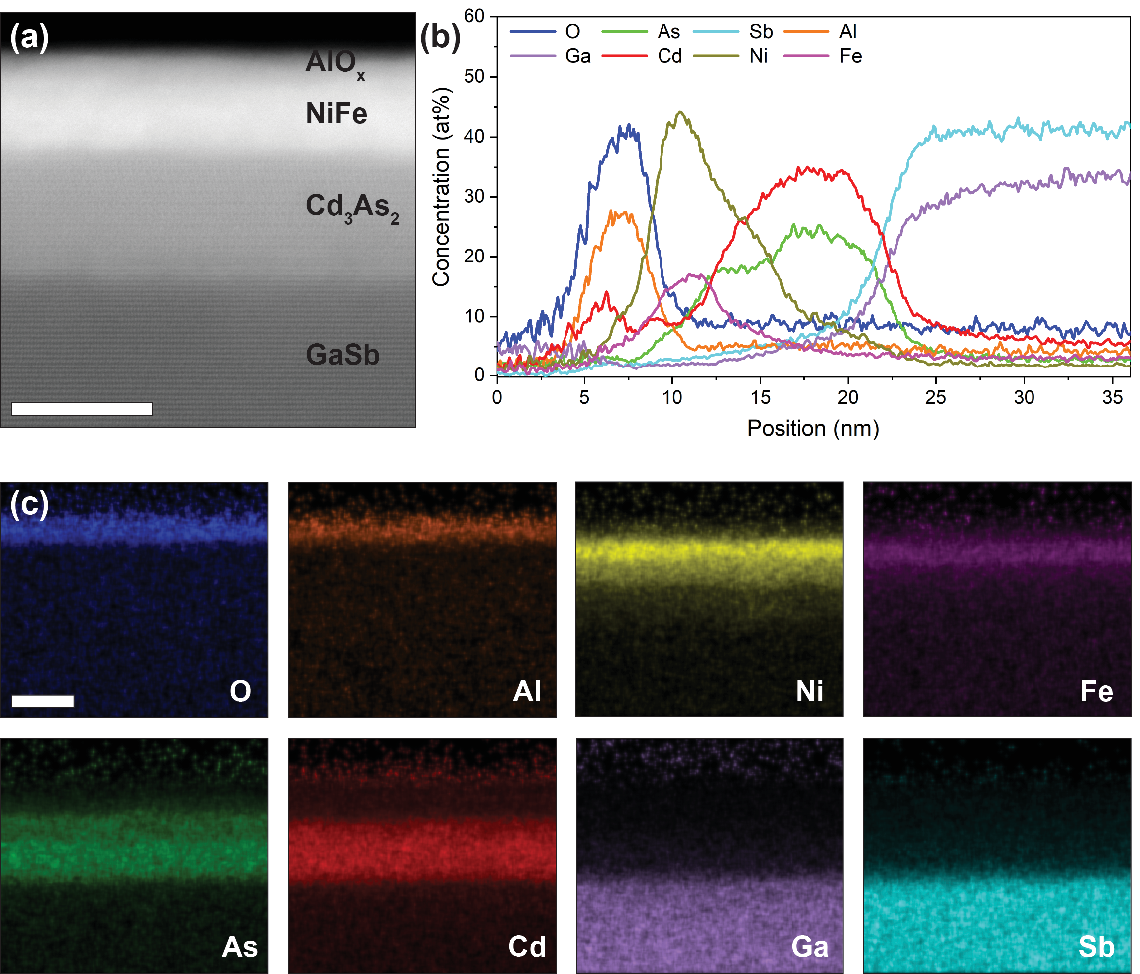}\\
\end{center}

\noindent {\bf Fig. 9.} (a) HAADF-STEM image of the top layers of the device. The scale bar is 10 nm.
(b) Concentration of the elements in (a) across the device interface layers shown. (c) Elemental maps of O, Al, Ni, Fe, As, Cd, Ga and Sb. Scale bar is 10 nm.

\begin{center}
\includegraphics[width=11cm]{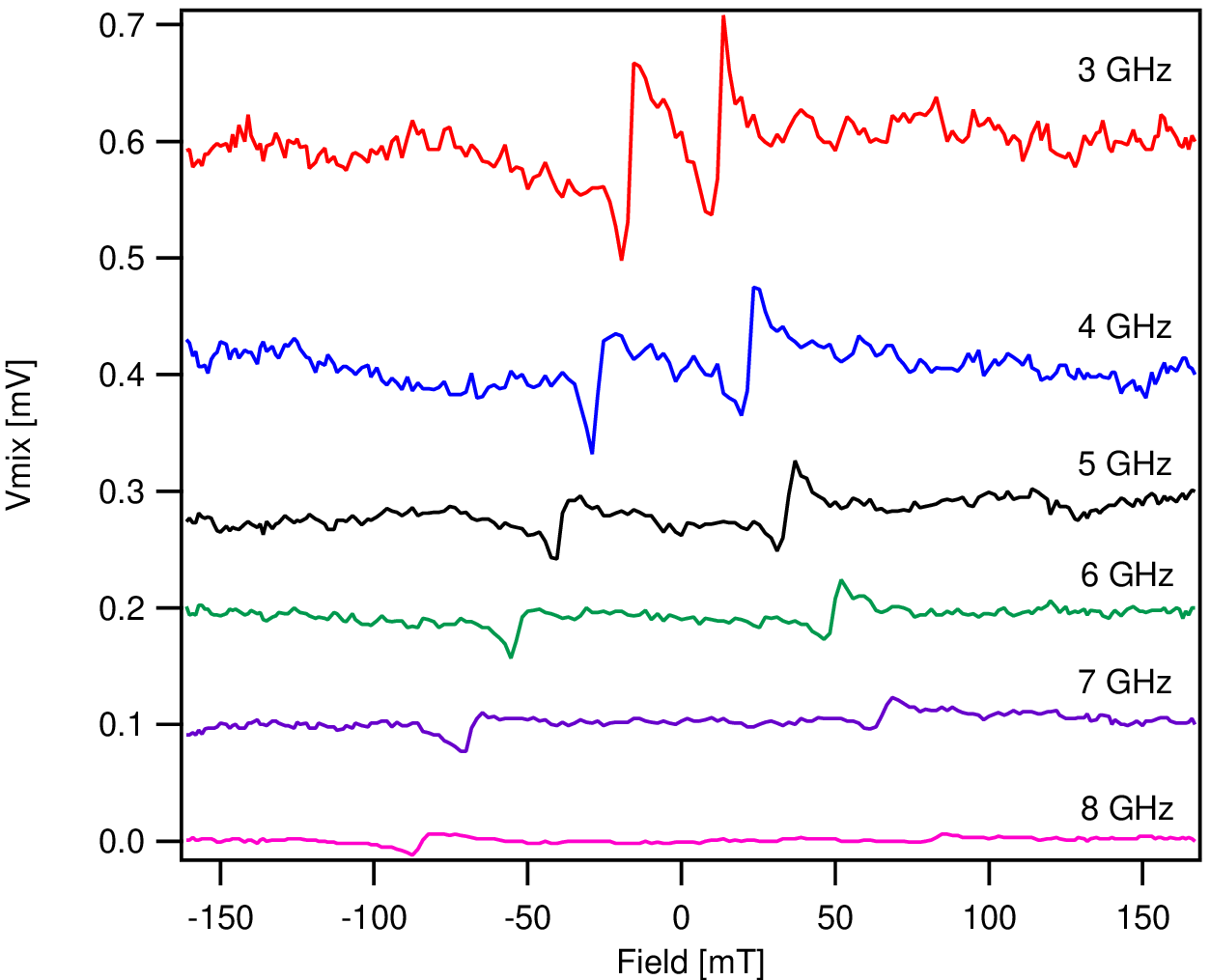}\\
\end{center}

\noindent {\bf Fig. 10.} ST-FMR mixing voltage measured at different frequencies in a completely $\it{in vacuo}$ grown heterostructure. 

\section{PNR measurements in $\mathrm{Cd_3As_2/Py}$ heterostructures}

To examine the possibility of proximity-induced magnetism in the $\mathrm{Cd_3As_2}$ layer caused by the adjacent ferromagnetic Py layer, we performed room-temperature polarized neutron reflectometry (PNR) measurements at the PBR instrument at the NIST Center for Neutron Research (NCNR). The incident neutrons ($\lambda = 4.75 \mathrm{Å}^{-1}$) were spin polarized parallel or antiparallel to the 2 T applied in-plane magnetic field, and the specular reflectivity was measured as a function of Q, the momentum transfer vector along the film normal direction. The spin-dependent neutron reflectivity is sensitive to the nuclear and magnetic scattering length density (SLD) depth profiles, so that depth-resolved information on the structure and magnetization may be extracted through fitting. Since the 2 T in-plane field is more than sufficient to saturate the Py magnetization, no in-plane component of the magnetization is expected to be perpendicular to the applied field. Therefore, only the non-spin flip reflectivity cross sections, sensitive to the in-plane magnetization parallel to the applied field, were collected while the spin-flip cross sections, sensitive only to the in-plane magnetization perpendicular to the applied field, were not. We reduced and analyzed the data using the Reductus and Refl1D software programs, respectively \cite{Maranville:po5131,KIRBY201244}. Parameter uncertainties were estimated using a Markov chain Monte Carlo (MCMC) method as implemented in the DREAM algorithm of the BUMPS python package.

The best fit to the data, shown in Figure 11, was generated using a model which allowed for a proximity-induced magnetization of varying thickness in the $\mathrm{Cd_3As_2}$ at the interface with the Py layer. However, the best fit to the data yields no induced magnetization in the $\mathrm{Cd_3As_2}$ at the interface, indicating a lack of proximity-induced magnetization within measurement sensitivity. Since the thickness and magnetization of a proximity-magnetized layer are often highly coupled parameters, uncertainty analysis was performed using a modified model in which the proximity-magnetized layer thickness was fixed at a typical value of 2 nm. In this case, DREAM analysis indicates an extremely small upper limit ($95\%$ confidence interval) of 10 kA/m. The model shown does not incorporate the $\mathrm{CdO_x}$ layer observed in STEM imaging, as modeling which incorporated this layer did not yield a notably different fit from those that did not. This will be understood given that the extremely thin 1 nm layer yields oscillations with a periodicity of approximately $6.28\ \mathrm{nm}^{-1}$ in Q, far outside the measured range. Further, the expected $\mathrm{CdO_x}$ thickness is the same order of magnitude as the fitted roughness at the $\mathrm{Cd_3As_2/Py}$ interface. Lastly, we note that the theoretical nuclear SLD ($4.065\ \times\ 10^{-4}\ \mathrm{nm}^{-2}$) of $\mathrm{CdO_x}$ is between that of $\mathrm{Cd_3As_2}$ and Py. Taken together, these factors suggest that the structural features associated with the $\mathrm{CdO_x}$ layer are likely to be extremely muted. Rather, we propose that the $\mathrm{CdO_x}$ layer will blend into the $\mathrm{Cd_3As_2}$ interface and will likely instead manifest in the SLD profile as a small magnetically dead layer in the Py at the $\mathrm{Cd_3As_2}$ interface. Indeed, incorporating such a magnetically dead layer in Py near the $\mathrm{Cd_3As_2}$ interface into the model yields a thickness of 1.04 nm $\pm$ 0.07 nm, in excellent agreement with the STEM and EELS. PNR measurements therefore indicate a lack of proximity-induced magnetization at room-temperature and confirm the presence of a nonmagnetic oxide-like layer at the interface.

\begin{center}
\includegraphics[width=12cm]{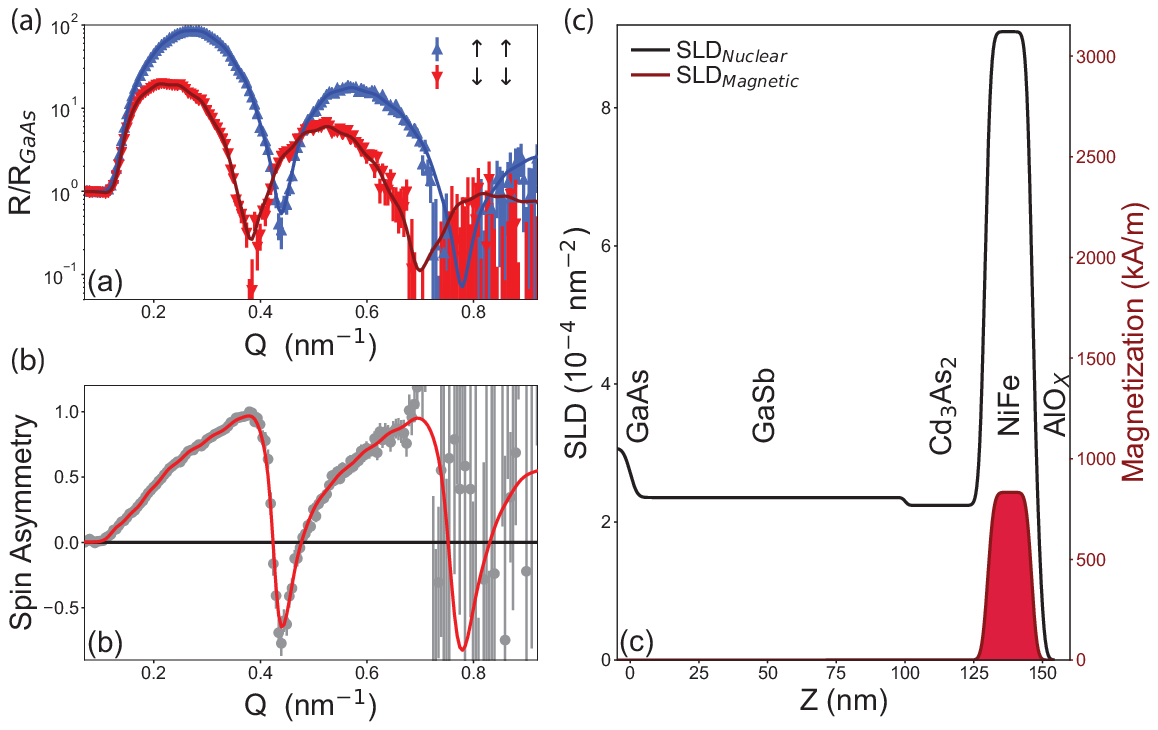}\\
\end{center}

\noindent {\bf Fig. 11} (a) Spin-dependent PNR data from a $\mathrm{Cd_3As_2/Py}$ bilayer at room-temperature in an applied field of 2 T alongside a theoretical fit. We show the Fresnel reflectivity, where the measured reflectivity is normalized by the Q-dependent theoretical reflectivity of the GaAs substrate. (b) Measured spin asymmetry, defined as (R$^{\uparrow\uparrow}$ - R$^{\downarrow\downarrow}$)/(R$^{\uparrow\uparrow}$ + R$^{\downarrow\downarrow}$) alongside theoretical fit. (c) Best fit nuclear and magnetic scattering length density (SLD) profiles used to generate the fits shown. The right axis shows the magnetization which is equivalent to a given magnetic SLD value.



\section{Summary of the samples measured in this study}

As mentioned in the main text, we have measured spin and charge interconversion in 12 devices. Table III shows the list of the samples used in this study and the measurements performed.
\begin{table} [b]
\caption{Samples used in this study indicating the nominal thickness of the $\mathrm{Cd_3As_2}$ and Py layers and the measurement performed to study the spin-charge interconversion. Samples with an * were used to calibrate the deposition rate via transmision electron microscopy and x-ray reflectivity (XRR).}
\begin{tabular}{lccc}
\hline
\makecell{Sample \\ number} & \makecell{$\mathrm{Cd_3As_2}$ \\ thickness (nm)} & \makecell{Py \\ thickness (nm)} & \makecell{Measurement \\ performed} \\ \midrule\midrule
190130A       & 40                & 30                & SP                    \\
190509A*       & 50                & 30                & SP, TEM, XRR               \\
190510A*       & 125               & 30                & SP, TEM, XRR               \\
190612A       & 50                & 30                & SP                    \\
190703A       & 75                & 30                & SP                    \\
190718A       & 150               & 30                & SP                    \\
190719A       & 200               & 30                & SP                    \\
190814A       & 80                & 4                 & ST-FMR                \\
190929A       & 20                & 4                 & ST-FMR,                \\
191002B*       & 19                & 4                 & ST-FMR, XRR                \\
191127B*       & 12                & 4, 5, 6           & SP, ST-FMR, XRR          \\
201125A*       & 8                 & 7                 & ST-FMR, TEM           \\ 
\hline
\end{tabular}
\label{Table3}
\end{table}

\newpage


\providecommand{\noopsort}[1]{}\providecommand{\singleletter}[1]{#1}%

\end{document}